\documentclass[conference]{IEEEtran}
\IEEEoverridecommandlockouts
\usepackage{cite}
\usepackage{amsmath,amssymb,amsfonts}
\usepackage{algorithmic}
\usepackage{graphicx}
\usepackage{textcomp}
\usepackage{xcolor}
\usepackage{comment}
\usepackage{todonotes}
\usepackage[caption=false]{subfig}
\usepackage{url}

\def\BibTeX{{\rm B\kern-.05em{\sc i\kern-.025em b}\kern-.08em
    T\kern-.1667em\lower.7ex\hbox{E}\kern-.125emX}}
\begin{document}

\title{PAGAN: Video Affect Annotation Made Easy}

\author{\IEEEauthorblockN{David Melhart}
\IEEEauthorblockA{\textit{Institute of Digital Games}\\
\textit{University of Malta}\\
Msida, Malta\\
david.melhart@um.edu.mt}
\and
\IEEEauthorblockN{Antonios Liapis}
\IEEEauthorblockA{\textit{Institute of Digital Games}\\
\textit{University of Malta}\\
Msida, Malta\\
antonios.liapis@um.edu.mt}
\and
\IEEEauthorblockN{Georgios N. Yannakakis}
\IEEEauthorblockA{\textit{Institute of Digital Games}\\
\textit{University of Malta}\\
Msida, Malta\\
georgios.yannakakis@um.edu.mt}
}

\maketitle
\begin{abstract}
How could we gather affect annotations in a rapid, unobtrusive, and accessible fashion? How could we still make sure that these annotations are reliable enough for data-hungry affect modelling methods? This paper addresses these questions by introducing PAGAN, an accessible, general-purpose, online platform for crowdsourcing affect labels in videos. The design of PAGAN overcomes the accessibility limitations of existing annotation tools, which often require advanced technical skills or even the on-site involvement of the researcher. Such limitations often yield affective corpora that are restricted in size, scope and use, as the applicability of modern data-demanding machine learning methods is rather limited. The description of PAGAN is accompanied by an exploratory study which compares the reliability of three continuous annotation tools currently supported by the platform. Our key results reveal higher inter-rater agreement when annotation traces are processed in a relative manner and collected via unbounded labelling.

\end{abstract}

\begin{IEEEkeywords}
Affective computing, human-computer interaction, affect annotation, crowdsourcing, video
\end{IEEEkeywords}

\section{Introduction}
The question of how to best collect many reliable and valid affect annotations is getting increasingly important in affective computing. Video-based annotation---as a popular approach in affective computing---requires participants to watch a set of videos and annotate their content, which is a cumbersome and costly process. Given the ever increasing use of data-hungry affect modelling techniques, however, the need for larger and more reliable affective corpora is growing. 




Meanwhile, the majority of current frameworks for both discrete and continuous video affect annotation pose a number of limitations. Tools such as FeelTrace \cite{cowie2000feeltrace}, ANNEMO \cite{ringeval2013introducing}, AffectButton \cite{broekens2013affectbutton}, GTrace \cite{cowie2013gtrace}, CARMA \cite{girard2014carma}, AffectRank \cite{yannakakis2015grounding}, and RankTrace \cite{lopes2017ranktrace} 
often require local installation and calibration. Not only does this require the presence of a researcher while conducting the study, but it often necessitates knowledge of a programming language as well. This low level of accessibility 
constrains the widespread use of such annotation tools (i.e. in the wild) and, in turn, results in affective datasets of limited size and use. This limitation is even more severe for newly emerging fields, such as game user research, without large established corpora, where data collection is a necessity. 



%

To address the above limitations this paper introduces a \emph{general-purpose}, \emph{online} video annotation platform, namely the \emph{Platform for Audiovisual General-purpose ANnotation} (PAGAN)\footnote{\url{http://pagan.institutedigitalgames.com/}}. The platform is publicly available and free to use for research purposes. PAGAN provides researchers with an easy and \emph{accessible} way to crowdsource affect annotations for their videos. In contrast to other popular annotation tools, PAGAN does not require a local installation and is designed to help researchers organise and disseminate their research projects to a large pool of participants. Inspired by \cite{mcduff2013affectiva}, the whole annotation process is done through a web interface operating on any modern web browser. Outsourcing the labelling task is as simple as sharing the corresponding project link.

PAGAN currently features three one-dimensional affect labelling techniques representing different methods for measuring the ground truth of affect: \emph{GTrace} \cite{cowie2013gtrace}, \emph{BTrace} (a modified version of AffectRank \cite{yannakakis2015grounding}) and \emph{RankTrace} \cite{lopes2017ranktrace}. In addition to the detailed description of the platform, the paper offers an exploratory study which examines the reliability of the three annotation methods. The results of this study reveal higher degrees of inter-rater agreement when traces are processed in a relative manner and collected via unbounded labelling. 



\section{Background}\label{sec:background}

PAGAN is centred on \emph{dimensional} and, primarily, \emph{continuous} affect annotation. To motivate this focus, this section presents the theoretical background of categorical versus dimensional emotional representations, and time-discrete versus time-continuous affect annotation techniques.

\subsection{Categorical vs. Dimensional Representation of Emotions}

Theories of emotions are generally represented in two main ways: as \emph{dimensions} or as \emph{categories}. The former focuses on emotions as emerging sentiments, which are functions of simple affective dimensions \cite{schlosberg1954three,mehrabian1980basic,russell1980circumplex}. The latter promotes an understanding, in which basic emotions are distinct from one-another in function and manifestation \cite{ekman1992argument,lazarus1996passion}. Today, both schools of thought have contemporary continuation, with some frameworks aiming to reconcile the two viewpoints \cite{cambria2012hourglass}. 

Categorical emotion representation is largely inspired by the work of Ekman \cite{ekman1992argument} and is based on the assumption that humans elicit distinct emotions, which are inherent to the human psyche and universally understood.
While normative studies have confirmed the generality of these frameworks to an extent \cite{diehl2007ekman,westbury2015avoid}, putting these theories into practice also brings about some conceptual limitations. The underlying assumption of clear division between basic emotional responses is challenged by a criterion bias when categorising fuzzy responses, and the subjective evaluation of emotions based on contextual cues highlights the relative nature of emotional appraisal \cite{aviezer2008angry} and calls the universality of these frameworks into question.

Alternatively, emotions can be represented through affective dimensions which typically follow Russell's \emph{Circumplex  Model of Emotions} \cite{russell1980circumplex} or the \emph{Pleasure-Arousal-Dominance model} \cite{mehrabian1980basic}. Many contemporary annotation tools \cite{morris1995sam,cowie2000feeltrace,cowie2013gtrace,lopes2017ranktrace}
use one of these models for annotating one or more affective dimensions.
The main limitation of these frameworks is that they cannot describe complex and self-reflexive emotions without expert interpretation, which could reintroduce biases to the observations. However, this simplicity also results in high \emph{face validity} \cite{nevo1985face}, reducing guesswork and \emph{criterion bias} of the annotator \cite{martinez2014don} even in the case of fuzzy responses, which otherwise would be hard to categorise.



PAGAN focuses on time-continuous annotation to capture the temporal dynamics of affective experiences. Because this task often involves identifying fuzzy transitions between affective responses, it relies on a dimensional representation of emotion. The choice of this focus is also motivated by the relatively low cognitive load of one-dimensional labelling compared to evaluating the manifestations and transitions of multiple distinct emotional categories.

\subsection{Discrete vs. Continuous Annotation Methods}

Traditional surveys, such as the Self-Assessment Manikin \cite{morris1995sam}, were developed to measure fixed scales with discrete items. While computer interfaces allowed for the development of time-continuous annotation tools, traditional surveys and digital tools for discrete affect annotation \cite{broekens2013affectbutton,yannakakis2015grounding} are still prevalent. Although these approaches capture less of the temporal dynamics of the experience \cite{metallinou2013annotation}, compartmentalising annotations could help reduce the noise of the labels and yield higher inter-rater agreement. Yannakakis and Martinez compared the nominal and ordinal representation of discrete affect annotations (AffectRank) with continuous bounded ratings (FeelTrace) \cite{yannakakis2015grounding}. Their study found that a nominal representation yields higher inter-rater agreement compared to treating a continuous trace as interval data. 

Treating continuous annotation traces as interval data and processing them in an absolute fashion remains the prominent method of many studies \cite{gunes2013categorical,metallinou2013annotation}. As this methodology necessitates that the trace is bounded to ensure a common scale among raters, interval processing of annotation traces provides data in a form that can be analysed via a wide array of statistical and machine learning approaches. However, there are serious caveats in representing inherently subjective experiences in an absolute fashion. Supported by the \emph{adaptation level theory} \cite{helson1964adaptation}, \emph{habituation} \cite{solomon1974opponent}, the \emph{somatic-marker hypothesis} \cite{damasio1994descartes}, and numerous studies within affective computing \cite{yannakakis2017ordinal,yannakakis2018ordinal} it appears that subjects experience stimuli in relation to their prior emotional and physiological states, experiences, and memories. Thus, any annotation task is subject to a number of \emph{anchoring} \cite{seymour2008anchors}, \emph{framing} \cite{tversky1981framing} and \emph{recency} \cite{erk2003emotional} effects.

PAGAN aims to overcome the above limitations by featuring unbounded continuous annotation (via RankTrace). With this labelling protocol, the data is no longer structured along the same scale, which makes processing traces as absolute values problematic. However, this method overcomes the limitations of interval processing that arise from the discrepancy between the players' cognitive evaluation processes and absolute scales. The relative processing of unbounded annotation traces has been shown to correlate with physiological signals \cite{lopes2017ranktrace}, and predictive models based on these traces have been shown to generalise better \cite{camilleri2017towards}.

\subsection{Annotation Tools}

In recent years, tools for affect labelling have diversified. Earlier examples such as FeelTrace  \cite{cowie2000feeltrace} in 2000, AffectButton \cite{broekens2013affectbutton} in 2009, and AffectRank \cite{yannakakis2015grounding} in 2014 aim to capture a complex phenomenon by measuring two or three affective dimensions at once. Yet recent studies \cite{cowie2013gtrace,girard2014carma,lopes2017ranktrace} focus on one-dimensional labelling. The shift away from multi-dimensional labelling can be explained by the increased cognitive load induced by these methods that comes with more complex tasks. Increased cognitive load can undermine the strengths of dimensional emotion representation \cite{cowie2013gtrace}, as one emotional axis can take precedence over the other---which could impact face validity. PAGAN implements three variations of one-dimensional affect labelling techniques, representing different methods for measuring the ground truth of affect: \emph{GTrace}, as bounded and continuous; \emph{BTrace} (binary trace), as real-time discrete; and \emph{RankTrace}, as unbounded and continuous annotation techniques.

GTrace \cite{cowie2013gtrace} was created as a bounded, continuous annotation tool and quickly became popular for affect labelling in human-computer interaction and affective computing \cite{baveye2015deep,muller2015emotion,dellandrea2016mediaeval,dhamija2018automated}. GTrace has a limited memory and displays only the last few annotation values. PAGAN implements its own version of GTrace based on the description of the tool in \cite{cowie2013gtrace} and \cite{baveye2015deep}.

Due to concerns regarding traces that are processed as intervals, \emph{AffectRank} \cite{yannakakis2015grounding} was introduced as a real-time rank-based discrete labelling tool. As a two-dimensional annotation tool, the main improvement of AffectRank over FeelTrace was the focus on recording ordinal changes instead of absolute values. BTrace in PAGAN is inspired by AffectRank as it measures ordinal change, but is much simpler: instead of the 8 annotation options in AffectRank, BTrace focuses on one affective dimension and two nominal labels: positive vs. negative change. A major limitation of both AffectRank and BTrace, however, is the discrete nature of the provided labels which limits the resolution of the collected ground truth data.

\begin{figure}[!tb]
    \centering
    \includegraphics[width=1\linewidth]{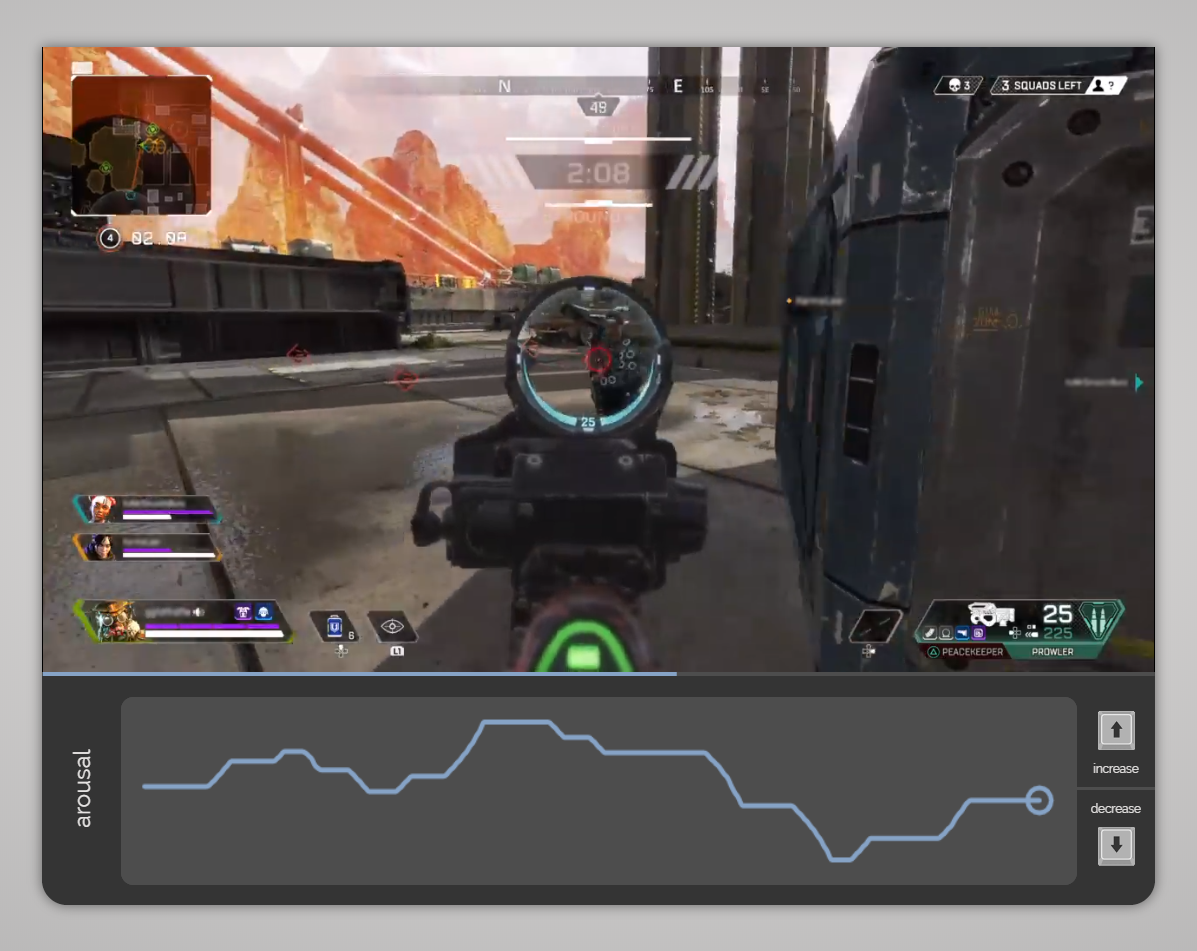}
    \caption{RankTrace interface in the PAGAN platform. \emph{Apex Legends} (Electronic Arts, 2019) gameplay footage. No copyright infringement intended.} \label{fig:ranktrace}
\end{figure}

To cater for the subjects' relative judgement models, \emph{RankTrace} was introduced for unbounded and relative annotation \cite{lopes2017ranktrace}. In RankTrace the annotator effectively draws a graph of their experience (Fig. \ref{fig:ranktrace}) which acts as the annotators' point of reference. 
RankTrace produces continuous and unbounded traces which can be observed as ordinal changes and be processed in a relative manner \cite{lopes2017ranktrace,camilleri2017towards}. Because RankTrace is unbounded, it lets the annotator react to the situation compared to previous experiences instead of forcing them to evaluate the stimuli in an absolute manner. In addition to GTrace and BTrace, the current PAGAN framework features a version of the RankTrace annotation method.

\section{PAGAN Platform Description} \label{sec:pagan}

\begin{figure}[!tb]
    \centering
    \includegraphics[trim={0cm 8.5cm 0cm 0cm}, clip, width=1\linewidth]{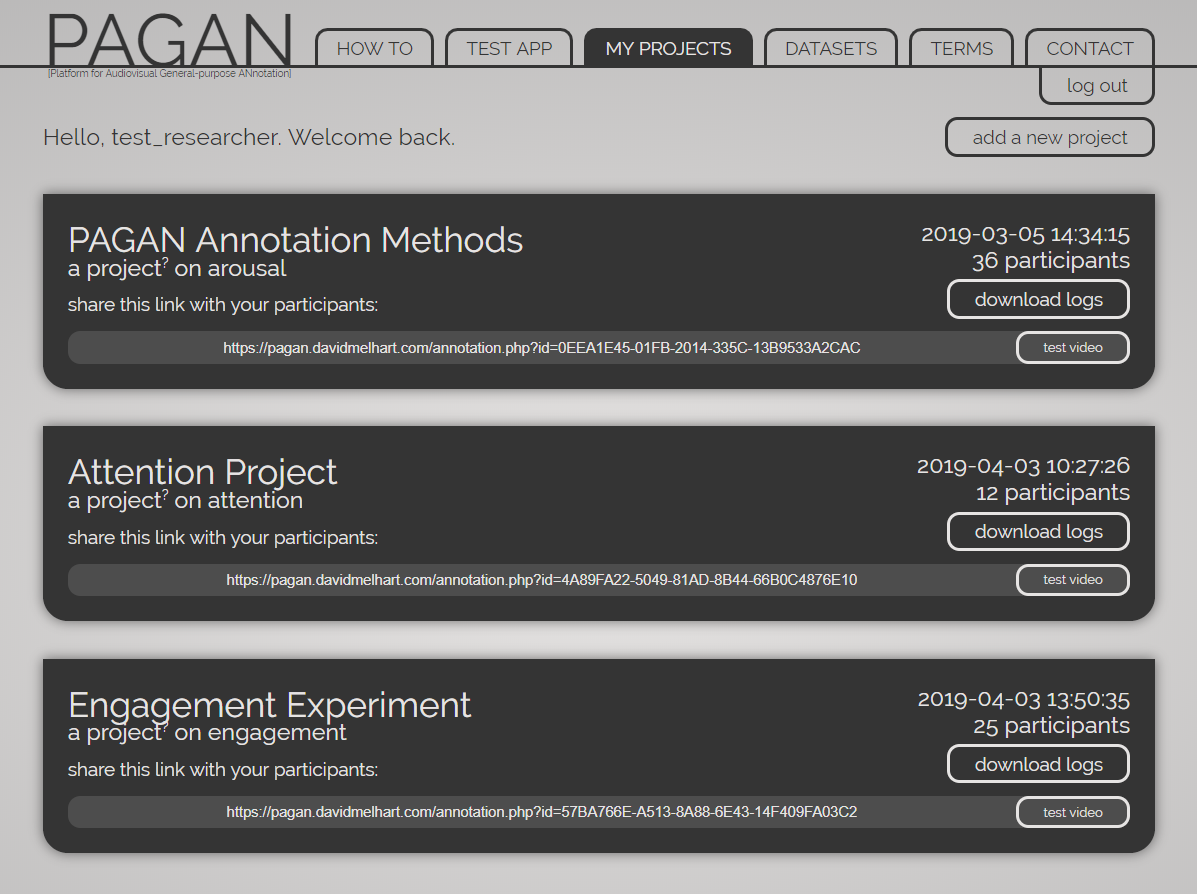}
    \caption{Project summaries on the administration interface.} \label{fig:projects}
\end{figure}

This section provides a description of the PAGAN platform, its user interface and general usage. The user interface of PAGAN consists of two separate sections. One is a web interface for researchers to prepare the annotation task (Section \ref{sec:researcher}) and the other is an interface for annotation by end-users (Section \ref{sec:annotator}). Section \ref{sec:annotation-methods} details the three annotation methods incorporated currently in PAGAN and used in the evaluation study of Section \ref{sec:study}.

\subsection{Administration Interface} \label{sec:researcher}

\begin{figure}[!tb]
    \centering
    \includegraphics[width=1\linewidth]{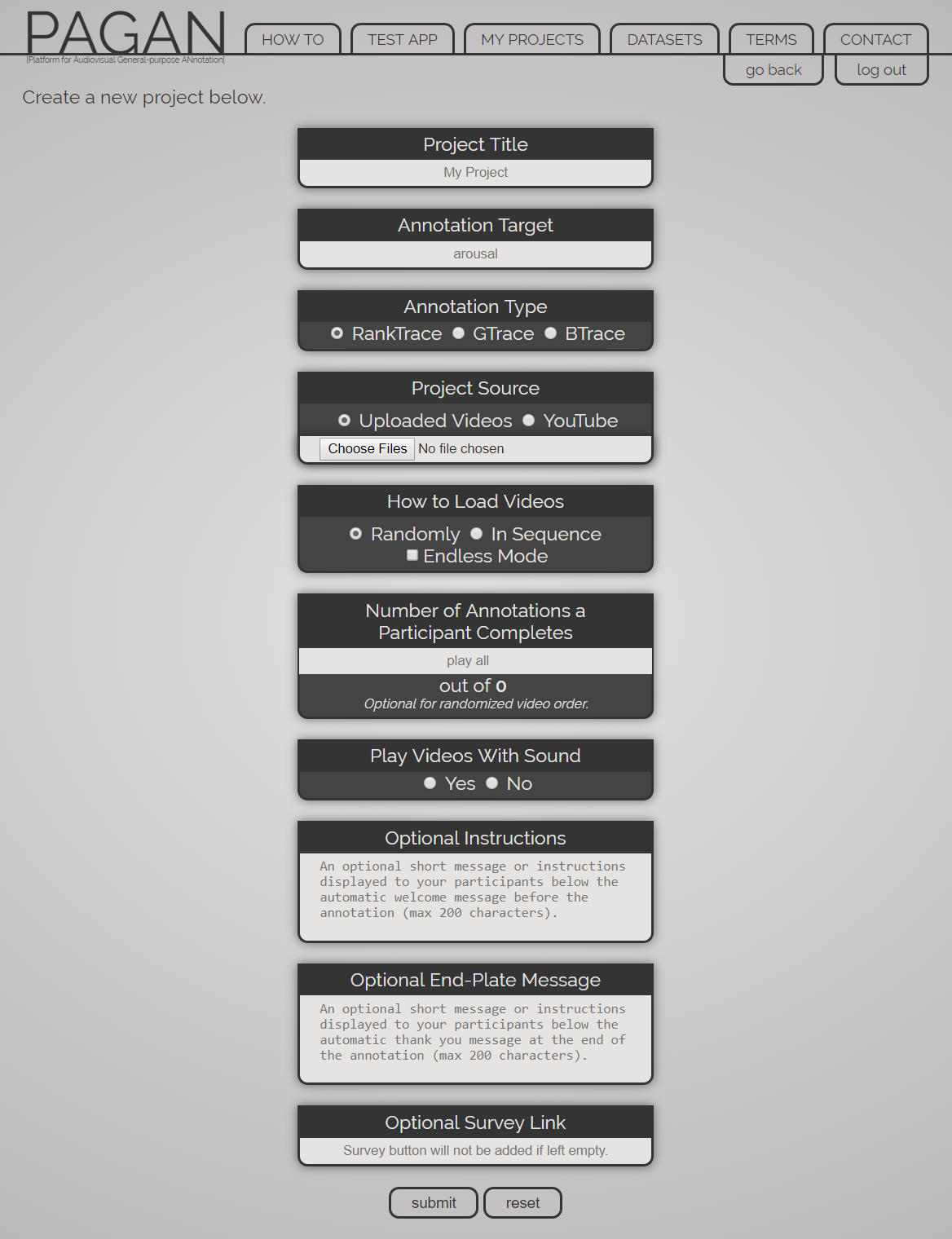}
    \caption{Project creation screen on the researcher interface.} \label{fig:create}
\end{figure}

Researchers access and create their projects through a dedicated page. Each user has a secure login with a username and a password. After login, the researcher accesses their \emph{project summaries} (Fig. \ref{fig:projects}). Here, they can create new projects, view the progress of their ongoing studies, and access their corresponding annotation logs. Each project has a corresponding link, which is meant to be shared with potential participants. The annotation application can also be run in \emph{test mode} from here, in which case the annotation logs are not saved.

The project creation screen can be seen in Fig. \ref{fig:create}. Projects are highly customisable to accommodate different research needs. 
The project title identifies the study on the project summary page and displayed to the participants as part of the welcome message. The \emph{annotation target} is the label for the $y$ axis of the annotator (see Fig. \ref{fig:ranktrace}). The project can be sourced from one or more uploaded videos or YouTube\footnote{\url{https://www.youtube.com/}} links. The videos can be loaded either randomly or in sequence. If \emph{endless mode} is selected, PAGAN rotates the videos indefinitely, allowing a participant to complete all tasks multiple times. In case of a randomised video order, there is an option to limit the number of videos a participant has to annotate. The videos can be played with or without sound; if videos are played with sound, PAGAN reminds participants to turn on their speakers or headphones. The researcher can optionally add information or instructions viewed before and/or after the annotation tasks to help integrate the platform into the larger research project. Finally, a survey link can be included, which is displayed to the participant at the end of the annotation session.

\subsection{Annotator Interface}\label{sec:annotator}

\begin{figure}[!tb]
    \centering
    \includegraphics[width=1\linewidth]{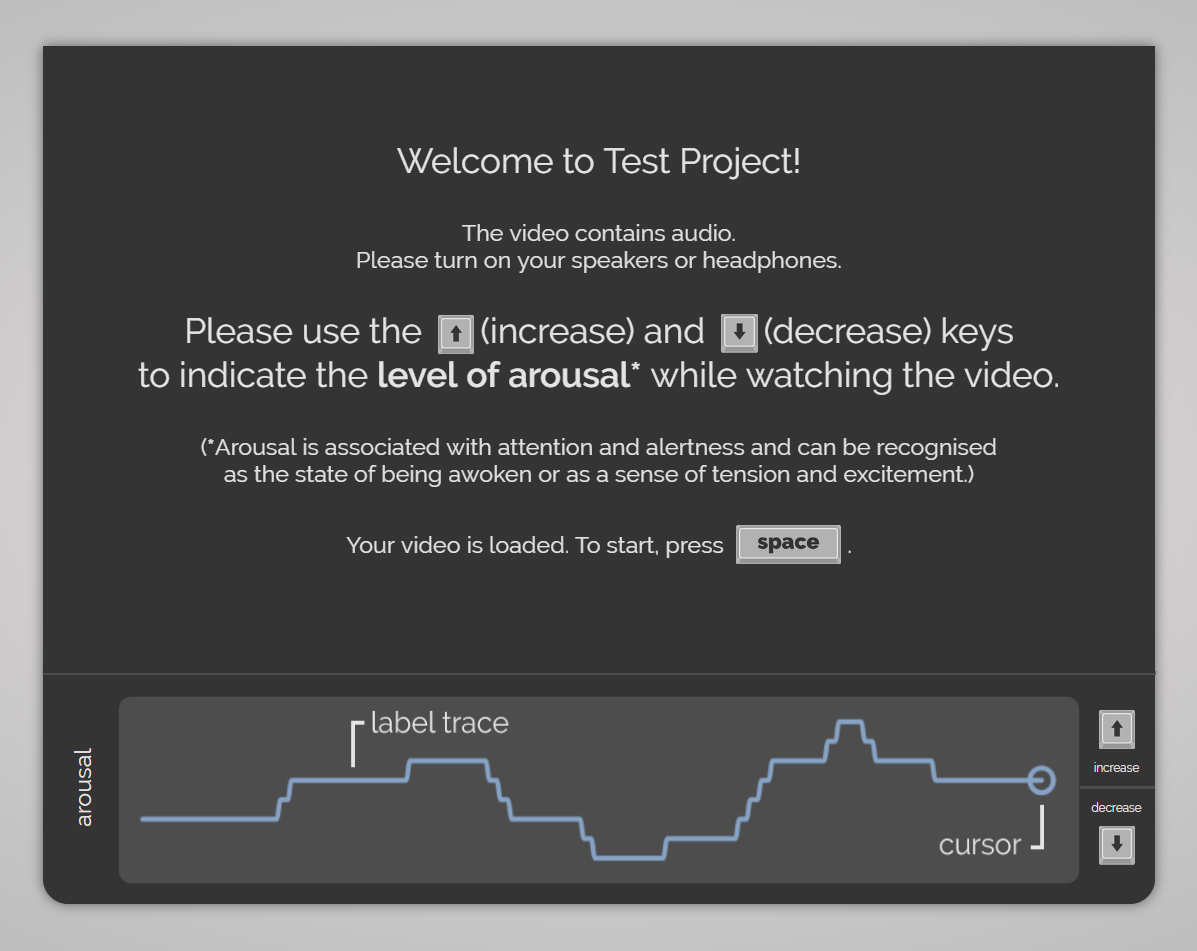}
    \caption{Welcome message displayed to participants before annotation starts.} \label{fig:welcome}
\end{figure}

The annotator application is a separate interface from the researcher site and meant to be used by the participants of the study. The interface is designed to display only the necessary information, thus eliminating potential distractions. Upon navigating to the project link (see Section \ref{sec:researcher}), the participant is greeted by a welcome message which concisely explains the annotation procedure and provides some information about the annotation target (Fig. \ref{fig:welcome}). 
After the video is loaded, the participant can start the annotation process (Fig. \ref{fig:ranktrace}) at their leisure. The design of PAGAN eliminates the use of a computer mouse in favour of the more readily-available keyboard. The annotation is performed with the \emph{up} and \emph{down} keys on the keyboard and the session can be paused by pressing \emph{space}. To minimise the amount of sessions with insufficient annotation, the system only logs a session as ``completed'' if at least $25\%$ is seen and pauses if the browser tab is out of focus (i.e.~if the participant leaves the annotation interface open but switches to a different tab or window). 

\subsection{Annotation Methods}\label{sec:annotation-methods}

This section presents the annotation techniques included in the PAGAN framework: RankTrace, GTrace, and BTrace.

\subsubsection{RankTrace} 

The implementation of RankTrace closely follows the original by Lopes et al. \cite{lopes2017ranktrace} (Fig. \ref{fig:ranktrace-ui}). The only major distinction to their version is the exponential acceleration of the annotator cursor when a control key is held down. This change was made because the original version of RankTrace uses a wheel interface where the magnitude of change can be controlled easier by the participant. As the annotation trace displays the entire history, the participant has sufficient visual feedback which acts as a reference (anchoring) point \cite{yannakakis2018ordinal} for the subjective evaluation of the experience.

\subsubsection{GTrace} 

Similarly to how the tool is used in Baveye et al. \cite{baveye2015deep}, the user interface is moved under the video; vertical lines are added as an allusion to a traditional 7-item scale to provide a visual aid for the absolute evaluation of the trace (Fig. \ref{fig:gtrace-ui}). Similarly to RankTrace, the movement of the cursor is accelerated when a key is held down as the original implementation used a mouse cursor allowing for higher speed while retaining precision. When the participant stops the cursor, it leaves a mark which slowly fades, providing limited memory of previous positions, to which the participant can compare new labels. The limited memory from the fading mark differs from both BTrace and RankTrace which display the full history of the session.

\subsubsection{BTrace} 

\begin{figure}[!tb]
    \centering
    \subfloat[RankTrace\label{fig:ranktrace-ui}]{
    \includegraphics[width=0.98\linewidth]{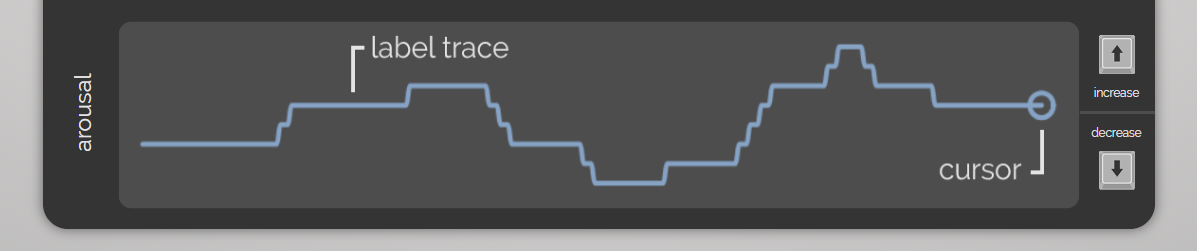}}
    \quad
    \subfloat[GTrace\label{fig:gtrace-ui}]{
    \includegraphics[width=0.98\linewidth]{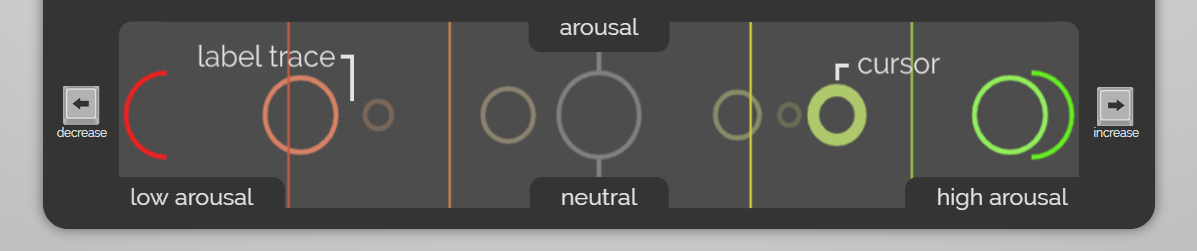}}
    \quad
    \subfloat[BTrace\label{fig:binary-ui}]{
    \includegraphics[width=0.98\linewidth]{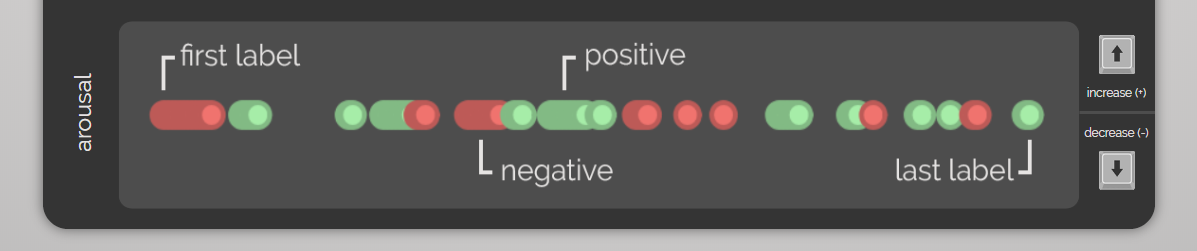}}
    \caption{Interfaces for the annotation methods included in this study.} \label{fig:ui}
\end{figure}

Binary Trace (BTrace) is a new annotation tool introduced in this paper which is largely based on AffectRank \cite{yannakakis2015grounding}. BTrace is designed as a simple alternative to relative annotation in a discrete manner, using two nominal categories: $+1$ as increase (or positive change) and $-1$ as decrease (or negative change). In that regard, it could be viewed as an one-dimensional version of AffectRank. The design of the tool, however, is based on the benefits reference points have on the reliability of the obtained annotation labels \cite{yannakakis2018ordinal,lopes2017ranktrace} and thus it displays the full history of the annotation session as red and green blobs (see Fig.~\ref{fig:binary-ui}).

\section{Example Study} \label{sec:study}

This section presents a small-scale exploratory study conducted with the PAGAN platform. The goal of this study is two-fold. First, we present the usage of the system in a real-world scenario; and second, we examine the effectiveness of relative annotation methods compared to absolute affect labelling. This study focuses on the perceived arousal level of different videos with emotional content. Our two relative annotation methods are RankTrace and Btrace, and our absolute method is a variant of GTrace (see Sec. \ref{sec:annotation-methods} and Fig. \ref{fig:ui}).

\subsection{Collected Data}

The collected data consists of $108$ annotated videos from $36$ participants. Participants were found through the social and academic network of the researchers, while subsequent parties were added through snowball sampling by participants sharing the project link. The average age of the participants is $29$ years old and $69\%$ identified as male, $24\%$ identified as female, one subject identified as queer and one did not want to identify themselves. The majority of the participants were avid gamers, with $59\%$ playing more than once a week. Each participant was asked to annotate three videos with different but emotionally evocative content: (a) recorded gameplay from \emph{Apex Legends} (Electronic Arts, 2019) (Apex), a popular Battle Royale-style game; (b) the Season 8 trailer of the TV series \emph{Game of Thrones} (HBO, 2019) (GoT); (c) a conversation between a human participant and ``Spike'', the angry virtual agent in the \emph{SEMAINE} database \cite{mckeown2012semaine}. All videos are approximately 2 minutes long. Each video was assigned a random annotation type, discussed in Section \ref{sec:annotation-methods}. The order of videos was also randomised. 

\begin{table}[!tb]
    \caption{Summary of participants' choice of most and least interesting video and most and least intuitive annotation tool}
    \label{tab:preference}
    \centering
    \begin{tabular}{c@{ }c@{}c|c@{ }c@{}c|c@{ }c@{}c}
    \hline\hline
    \multicolumn{9}{c}{\textbf{Most / Least Interesting Video}}\\
    \hline
\multicolumn{3}{c|}{Apex}& \multicolumn{3}{c|}{GoT} & \multicolumn{3}{c}{SEMAINE} \\
\hline
11&/&6 & 	15&/&1 & 	3&/&22 \\
\hline\hline 
    \multicolumn{9}{c}{\textbf{Most / Least Intuitive Tool}}\\
\hline		
\multicolumn{3}{c|}{RankTrace} & \multicolumn{3}{c|}{GTrace} & \multicolumn{3}{c}{BTrace} \\
\hline
16&/&4 & 	8&/&8 & 	5&/&17 \\\hline\hline
    \end{tabular}
\end{table}
\begin{table}[!tb]
    \centering
\caption{Number of annotation traces for each video and annotation type, the average number of samples are shown in brackets}
    \begin{tabular}{c || r r | r r | r r}
\textbf{Annotation Type}&\multicolumn{2}{c| }{\textbf{Apex}}&\multicolumn{2}{c| }{\textbf{GoT}}&\multicolumn{2}{c }{\textbf{SEMAINE}}\\
\hline
\hline
\textbf{RankTrace} & 11 & (229) & 10 & (514) & 10 & (425)\\
\textbf{GTrace} & 11 & (429) & 12 & (358) & 8 & (133)\\
\textbf{BTrace} & 9 & (302) & 8 & (120) & 14 & (85)\\
    \end{tabular}
    \label{tab:data-size}
\end{table}

Participants were asked to name the most and least interesting of the three videos and the most and least intuitive of the three annotation tools, effectively ranking them. The results of their preferences are summarised in Table \ref{tab:preference}. The GoT trailer was the most popular (only one participant rated it as the least interesting), while the video from the SEMAINE database was by far the least liked (it collected 81\% of ``least interesting'' votes). In terms of usability, participants ranked RankTrace the most intuitive (as it received $55\%$ of ``most intuitive'' votes), GTrace second, 
and BTrace the least intuitive.


\subsection{Methodology}

To measure the reliability of the different annotation techniques over the different videos, we observe the inter-rater agreement between participants. Inspired by Yannakakis and Martinez \cite{yannakakis2015grounding}, we measure the inter-rater agreement with the \emph{Krippendorff's $\alpha$} coefficient \cite{krippendorff2004reliability}, which is a robust metric of the degree of agreement corrected for chance between any number of observers and any type of data. 
Krippendorff's $\alpha = 1 - {D_o}/{D_e}$, where $D_e$ denotes the expected and $D_o$ the observed disagreements between annotations.
%
Krippendorff's $\alpha$ is adjusted to the level of measurement of the observations through the weighing of the expected and observed coincidences (see \cite{krippendorff2011computing} for a complete explanation). This robustness allows for a fair comparison between different annotation methods. Krippendorff's $\alpha$ has an upper bound of $1$, which indicates absolute agreement, while $0$ signifies no agreement or pure chance. At Krippendorff's $\alpha<0$, disagreements between annotators are systematic and go beyond chance-based levels.

To allow for a comparison between discrete and continuous annotation and smooth out some of the surface differences between individual traces, we compartmentalise the signals into equal length time-windows. This method of preprocessing is often used in affective computing to preprocess time-continuous signals \cite{yannakakis2015grounding,camilleri2017towards,lopes2017ranktrace}. We clean the dataset of traces which either had extremely few samples from annotation (less than 3) or where viewing time was less than a minute. This cleanup process removed 15\% of traces, and the final datasets comprise of $92$ traces. Table \ref{tab:data-size} shows the number of traces and samples in each dataset and annotation method. In this study $3$-second time windows are considered without any overlap. Potentially the $3$-second processing provides approximately $40$ 
windows per participant. As some participants did not complete the full annotation task, this number can vary. However, to maximise the sample sizes, we decided to keep these traces as Krippendorff's $\alpha$ can be applied to data with missing observations as well.

As BTrace already encodes perceived change, similarly to AffectRank \cite{yannakakis2015grounding}, we compute the value of time windows as the \emph{sum of annotation values} ($\Sigma_A$) within each window, adding values in case of increase and subtracting them in case of decrease. For RankTrace and Gtrace, we consider both an absolute and a relative metric \cite{camilleri2017towards}: the mean value ($\mu_A$) and average gradient ($\Delta_A$) of time-windows based on the min-max normalised traces. We consider the mean value an absolute metric because it denotes the general level of the participant's response in a given time-window. In contrast, the average gradient of a time-window considers the amount and direction of the change that happened, as it is computed from the differences of adjacent datapoints of the trace \cite{camilleri2017towards, lopes2017ranktrace}. The calculation of Krippendorff's $\alpha$ is adjusted to the observed metric. When the annotation trace is processed into a relative metric ($\Sigma_A$, $\Delta_A$), we compare annotation values as ordinal variables. When the annotation trace is processed into an absolute metric ($\mu_A$), we compare annotation values as interval variables.


\subsection{Results}

\begin{table}[!tb]
    \centering
\caption{Krippendorff's $\alpha$ across annotation traces processed as 3-second time windows} 
    \begin{tabular}{l c || r r r}
\multicolumn{2}{c ||}{\textbf{Annotation}}&\multicolumn{3}{c}{\textbf{Video}}\\
\hline
\hline
\textbf{Tool} & \textbf{Processing} & \textbf{Apex} & \textbf{GoT} & \textbf{SEMAINE}\\
\hline
\textbf{RankTrace} & $\Delta_A$ & 0.2025 & 0.1760 & -0.0043\\
 & $\mu_A$ & 0.1542 & -0.0227 & 0.0147\\
 \hline
\textbf{GTrace} & $\Delta_A$ & 0.1517 & 0.1224 & -0.0347\\
 & $\mu_A$ & 0.1857 & 0.0549 & 0.0926\\
 \hline
\textbf{BTrace} & $\Sigma_A$ & 0.3193 & 0.0973 & 0.0249\\
    \end{tabular}
    \label{tab:3sec-window}
\end{table}

This section presents the results of the statistical analysis and an interpretation of the results. The calculated inter-rater agreement based on Krippendorff's $\alpha$ scores are displayed in Table \ref{tab:3sec-window}. For the purpose of comparisons of RankTrace and GTrace, we use the highest $\alpha$ value between $\Delta_A$ and $\mu_A$.

The highest $\alpha$ values for RankTrace are 0.20 for Apex and 0.18 for GoT, which are higher than the highest $\alpha$ values for GTrace (0.19 and 0.12 respectively). For both GoT and Apex videos, the highest $\alpha$ values are found with $\Delta_A$ in three of the four instances examined (except for annotations with GTrace on the Apex dataset), which is further evidence that processing time-windows of GTrace ratings through a relative measure yields more consistent results.
Interestingly, both GTrace and RankTrace have a higher $\alpha$ value with $\mu_A$ for the SEMAINE video (with GTrace having superior performance), although generally these values are very low and any inter-rater agreement could be chance-based.
The general findings from these comparisons are in line with a growing body of research promoting the relative collection and processing of affective annotation traces \cite{yannakakis2015ratings,camilleri2017towards,yannakakis2018ordinal}.

Based on Table \ref{tab:3sec-window}, it seems that BTrace achieves the highest inter-rater agreement on the Apex dataset, while showing lower reliability on GoT and SEMAINE videos. As the compartmentalised binary labels denote the rough amount of perceived change in a time-window (but not its magnitude), the possibility of relatively high inter-rater agreement is not surprising. However, results on the GoT and SEMAINE videos show the unreliability of this method. A possible reason for the high variance in the inter-rater agreement is the low face validity of the method. BTrace collected 59\% of the ``least intuitive'' votes among the three annotation methods. Therefore, despite its potential robustness in certain cases, BTrace has shown to be the least reliable and intuitive to use.

An unexpected finding of this analysis is the overall low inter-rater agreement of all methods on the chosen SEMAINE video, which was also ranked as the least interesting by participants. A plausible explanation of the results is a connection between the context and intensity of the affective content and the reliability of the annotation traces. While games and trailers are designed to elicit arousal, the slow pace of SEMAINE videos can be unappealing by comparison.
The differences in inter-rater agreements between the Apex and GoT datasets also point towards the role of context in emotion elicitation. While the GoT video is authored to elicit high arousal, the Apex footage presents a more organic scenario with relative calm periods and high-octane action. Especially for frequent videogame players, who have personal experiences with the dynamics of shooter games, this video is easier to interpret and the affective high-points are easier to recognise. This is also supported by a recent study of Jaiswal et al. \cite{jaiswal2019muse}, who also observed an effect between the context of the annotation task and the quality of labels. 

\section{Discussion}\label{sec:discussion}
This paper presented an online platform for crowdsourcing affect annotations, 
providing researchers with an accessible tool for labelling any kind of audiovisual content.
A companion study showcased the usability of the platform, highlighted the reliability of the supported annotation techniques,
and compared  bounded, unbounded and binary annotations of arousal.
Results showed that an unbounded relative annotation method which includes the entire history of labels as reference points is more intuitive to use.
Moreover, the study included three videos indicative of different sources of arousal: a game, a TV series trailer, and a dialogue with a virtual agent. Our analysis reveals low inter-rater agreement on the SEMAINE database video, which raises the question on whether more engaging forms of emotion elicitation such as games would offer more reliable benchmarks for affective computing research.



The main limitation of the user study of Section \ref{sec:study} is the preliminary nature of the analysis. Arguably with a more thorough pruning, regularisation of the annotation traces \cite{wang2018towards}, quality control of the labels using gold-standards \cite{burmania2016increasing}, or a strict selection process for the included participants \cite{soleymani2010crowdsourcing}, higher inter-rater agreement can be achieved. Moreover, the exploratory nature of this study assessed how different types of videos can be annotated; in a more concise study the set of videos should likely be both larger and more consistent in terms of subject matter. However, as the main focus of this paper was the introduction of the PAGAN platform, these explorations were out of scope of the current study.

There is a large number of features that can be incorporated into the PAGAN platform, such as the support for more flexible research protocols through participant uploads. In the future, PAGAN can be extended with data preprocessing, analysis, and visualisation tools, providing researchers a toolbox for not just data collection but preliminary analysis as well. Such a toolbox could include automatic processing of traces into time windows, outlier detection and pruning; statistical summary and analysis in terms of inter-rater agreement. 
Machine learning support can be integrated with PAGAN as well, either to preprocess and format data for other software, such as the \emph{Preference Learning Toolbox}\cite{farrugia2015preference,camilleri2019py}
or as a light-weight predictive modelling module in PAGAN itself. 

\section{Conclusion}\label{sec:conclusion}
This paper presented a highly customisable and accessible online platform to aid affective computing researchers and practitioners in the crowdsourcing process of video annotation tasks. 
In a companion study, we demonstrated the reliability of the supported annotation techniques and showed the strength of relative annotation processing. 
Our key findings  advocate the use of relative, continuous, and unbounded annotation techniques and the use of videogames as active elicitors of emotional responses.

\section*{Acknowledgements}

This paper is funded, in part, by the H2020 project Com N Play Science (project no: 787476).

\bibliography{bibliography.bib}
\bibliographystyle{IEEEtran}
\end{document}